\documentclass[twocolumn,superscriptaddress,preprintnumbers,amsmath,amssymb,floatfix]{revtex4}
\usepackage{epsfig}
\usepackage{epstopdf}
\usepackage{graphicx}
\usepackage{grffile}
\usepackage{bm}
\usepackage{sidecap}
\usepackage{mathptmx}
\usepackage{txfonts}
\usepackage[usenames,dvipsnames,svgnames]{xcolor}
\usepackage{array}
\usepackage{float}
\usepackage{lipsum}
\newcommand\redout{\bgroup\markoverwith{\textcolor{red}{\rule[.5ex]{2pt}{0.4pt}}}\ULon}

\usepackage[colorlinks=true,citecolor=blue]{hyperref}
\hypersetup{colorlinks=true,citecolor=blue,linkcolor=blue,urlcolor=blue}
\begin{document}
\title{Strain impacts on commensurate bilayer graphene superlattices: distorted trigonal warping, emergence of bandgap and direct-indirect bandgap transition}
\date{\today}

\author{Zahra Khatibi}\email{za.khatibi@gmail.com}
\affiliation{Department of Physics, Iran University of Science and Technology, Narmak, 16846-13114, Tehran, Iran}
\author{Afshin Namiranian\footnote{Author to whom any correspondence should be addressed}}
\affiliation{Department of Physics, Iran University of Science and Technology, Narmak, 16846-13114, Tehran, Iran}
\author{Fariborz Parhizgar}
\affiliation{School of Physics, Institute for Research in Fundamental Sciences (IPM), Tehran 19395-5531, Iran}
\affiliation{Department of Physics and Astronomy, Uppsala University, Box 530, SE-751 21 Uppsala, Sweden}

\begin{abstract}
Due to low dimensionality, the controlled stacking of the graphene films and their electronic properties are susceptible to environmental changes including strain. The strain-induced modification of the electronic properties such as the emergence and modulation of bandgaps crucially depends on the stacking of the graphene films. However, to date, only the impact of strain on electronic properties of Bernal and AA-stacked bilayer graphene has been extensively investigated in theoretical studies. Exploiting density functional theory and tight-binding calculation, we investigate the impacts of in-plane strain on two different class of commensurate twisted bilayer graphene (TBG) which are even/odd under sublattice exchange (SE) parity. We find that the SE odd TBG remains gapless whereas the bandgap increases for the SE even TBG when applying equibiaxial tensile strain. Moreover, we observe that for extremely large mixed strains both investigated TBG superstructures demonstrate direct-indirect bandgap transition.
\end{abstract}
 \maketitle

\section{Introduction}\label{intro}
The stacking of graphene films adds an intriguing class of graphene-based 2D materials, with new and exceptional properties \cite{Mele2010,Zan2011,
Trambly2012,Kim2016a,cao2018unconventional}. This new class of materials that are recognized by the relative angle between the adjacent layers, namely moir\'e pattern, possess interesting angle-dependent properties which are different from that of bulk or monolayer graphene \cite{Shallcross2008,LopesdosSantos2012,McCann2013,Wang2013,Moon2013,San-Jose2013,Cosma2014,Chen2014,Moon2014a,Pham2014,Dai2016}. Added to the low-temperature superconductivity at magic twist angle \cite{cao2018unconventional}, the rotation dependent low-energy electronic behavior of the twisted bilayer graphene (TBG) includes fractional quantum hall effect \cite{Moon2012,Cao2016,Kim2017}, Van Hove singularities \cite{Li2009a,Brihuega2012}, the appearance of the secondary Dirac points \cite{Wallbank2013a,Mucha-Kruczynski2013,Nam2017}, the emergence of flat bands at the Fermi energy \cite{SuarezMorell2010,Luican2011a,Fang2016,Cao2018a}, and the reduction of group velocity in the limit of small twist angles leading to localization of Dirac electrons \cite{Bistritzer2010,Uchida2014}. The matching periodicity of the lower and upper layer of the TBGs forming the well established commensurate structures results in moir\'e pattern of longer periodicity than that of the Bernal and AA-stacked bilayer graphene (BG) reaching the high-wavelength of thousands of atomic distance \cite{LopesDosSantos2007,Latil2007,Xu2014,Wang2014b,Wong2015a,Koshino2015}. This group of TBGs are identified by the sublattice exchange (SE) parity and can be classified into two distinctive groups, odd and even, which resemble the low-energy characteristic of the Bernal and AA-stacked BG. The SE even structures are gapped due to pseudospin-orbit coupling, whereas the SE odd commensurate moir\'e structures have two massive bands which intersect at the charge neutrality point \cite{Mele2010}.

The low-energy electronic properties of stacking of graphene layers can be extensively affected by strain due to low dimentionality \cite{Mucha-Kruczynski2011,Savini2011,Frank2012,Amorim2015,Guisset2016}. Pseudoscalar potentials and transverse electric fields formed by different homogeneous in-plane strains, on each layer of BG, can lead to bandgap opening. For asymmetrically strained BG the bandgap is shown to undergo a transition from direct to indirect \cite{Choi2010a}. Out of plane strains can also lead to the formation of the bandgaps as a consequence of the enhancement of sublattice inequivalence when pulling the layers apart \cite{Wong2012}. Furthermore, compressive strain normal to the Bernal stacked BG results in the increment of the interlayer interactions leading to an enhancement in the Lifshitz transition \cite{Bhattacharyya2016}. For in-plane strains, on the other hand, results are found to be analogous to those of monolayer graphene, i.e. only expansion or compression along the zigzag direction can lead to the emergence of bandgaps \cite{Verberck2012}. For small angeled TBG, the energy separation of low-energy van Hove singularities is shown to decrease as the lattice deformation increases and well-defined pseudo-Landau levels emerge. Also, the joint effect of strain and out-of-plane deformation leads to valley polarization and formation of a significant gap \cite{Yan2013}.

Strain, however, is a costly method for tunning and emergence of the bandgaps for mono and specifically bilayer graphene \cite{Pereira2009a,Verberck2012} where a much higher interface shear stress is required compared to monolayer graphene for the same level of axial strain \cite{Frank2012} and thus the slippage of the layers on the substrate becomes inevitable \cite{Roldan2015}. Here, we show that the large angled commensurate TBGs are promising platform for manipulation of the electronic structure at low strain costs, especially because the fabrication of the moir\'e structures with controlled stacking is experimentally feasible \cite{Kim2016,Yankowitz2016a}. We investigate the electronic structure of different commensurate TBG superlattices with large misalignment angle close to Bernal stacked BG under in-plane strain. We aim to cover possible features driven by real space symmetries, specifically the SE parity and compare the electronic behavior of the two distinct SE odd and even structures when applying strain. To this end, we conduct DFT and tight-binding (TB) calculations and measure the strain-induced modification of the low-energy electronic structure. We find that when applying biaxial tensile strain, SE odd TBG superlattices remain gapless whereas the gap energy for the SE even TBG superstructures increases monotonically. We then take the advantage of the reasonably comparable results of {\it ab initio} and TB calculation to use the TB method as a less computationally expensive method to study the gap modulation for numerous strain configurations including small asymmetric strains ($<$5\%) where we observe direct-indirect bandgap crossover. 

The paper is organized as follows. In Sec.\ref{theo}, we discuss the details of the geometrical structure of commensurate superlattices and the methods we used. In Sec.\ref{res}, we demonstrate our results for the band dispersion of unstrained and strained TBGs, the modulation and relocation of band spacing, the behavior of the low-energy bands near the charge neutrality point and strained-induced changes of the gap energy for 441 different strain configurations. We summarize our findings in Sec.\ref{concl}.
\section{Theory}\label{theo}
TBGs consist of two graphene layers rotated by the angle $\theta$ with respect to each other around the vector perpendicular to their plane. Thus, the primitive vectors of the individual layers are related to each other as, $\vec{a}_{1(2)}^{\prime}=e^{i \theta}\vec{a}_{1(2)}$, where $\vec{a}_{1(2)}$ ($\vec{a}_{1(2)}^{\prime}$) is the primitive vector of the upper (lower) layer. Moreover, the lattice translation vectors of the upper and lower layers on the span of their primitive vectors can be written as  $\vec{\mathcal{T}}^{(\prime)}_{m^{(\prime)}n^{(\prime)}}=m^{(\prime)}~\vec{a}^{(\prime)}_1+n^{(\prime)}\vec{a}^{(\prime)}_2$, in which $m^{(\prime)}$ and $n^{(\prime)}$ are integers. The periods of the individual layers generally might not coincide with each other and hence the TBG structures become incommensurate. On the other hand, when at a specific angle $\theta_{mn}$ and distance $l$, the periods of the lower and upper layer coincide with each other, the commensuration takes place. In other words, while rotating the layers around the common fixed A sublattices at the origin, commensuration occurs when the translation vectors of the upper and lower layer addressing the next A sublattice become equal, i.e $\vec{\mathcal{T}}_{mn}=\vec{\mathcal{T}}^\prime_{m^\prime n^\prime}$ \cite{Moon2013}. Also, it can be shown that the total number of disclosed atoms in the commensurate supercell is $4(n^2+nm+m^2)$, and the relative rotation angle $\theta_{mn}$ at which commensuration takes place is \cite{Mele2012},
\begin{equation}
\theta_{mn}=\cos^{-1} \left({\frac{n^2+4nm+m^2}{2(n^2+nm+m^2)}}\right).
\end{equation}
Hereafter, we will use the notation $(m,n)$ to address commensurate superlattices throughout this study.

As discussed earlier, the low-energy electronic behavior of the moir\'e commensurate structures is strongly dependent on SE parity. Regarding SE parity, the commensurate structures generally can be addressed in two distinct groups of SE odd and even. A commensurate lattice is SE odd when only one sublattice site of the upper layer, (A) sublattice site, coincides with that of the lower layer. On the other hand, when two sublattice sites, (A) and (B), of the neighboring layers coincide, the commensurate moir\'e structure is even \cite{Mele2012}. While SE odd structures are gapless, the ones that are even under SE parity are gaped and have curved bands. Also, the low-energy behavior of the SE symmetric and asymmetric structures is resembling of their limiting cases, i.e. AA and Bernal stacked BG at $\theta=0$ and $\theta=\pi /3$, respectively \cite{Mele2012}.
%
%
%
%
\begin{figure}[t]
\includegraphics[width=\linewidth]{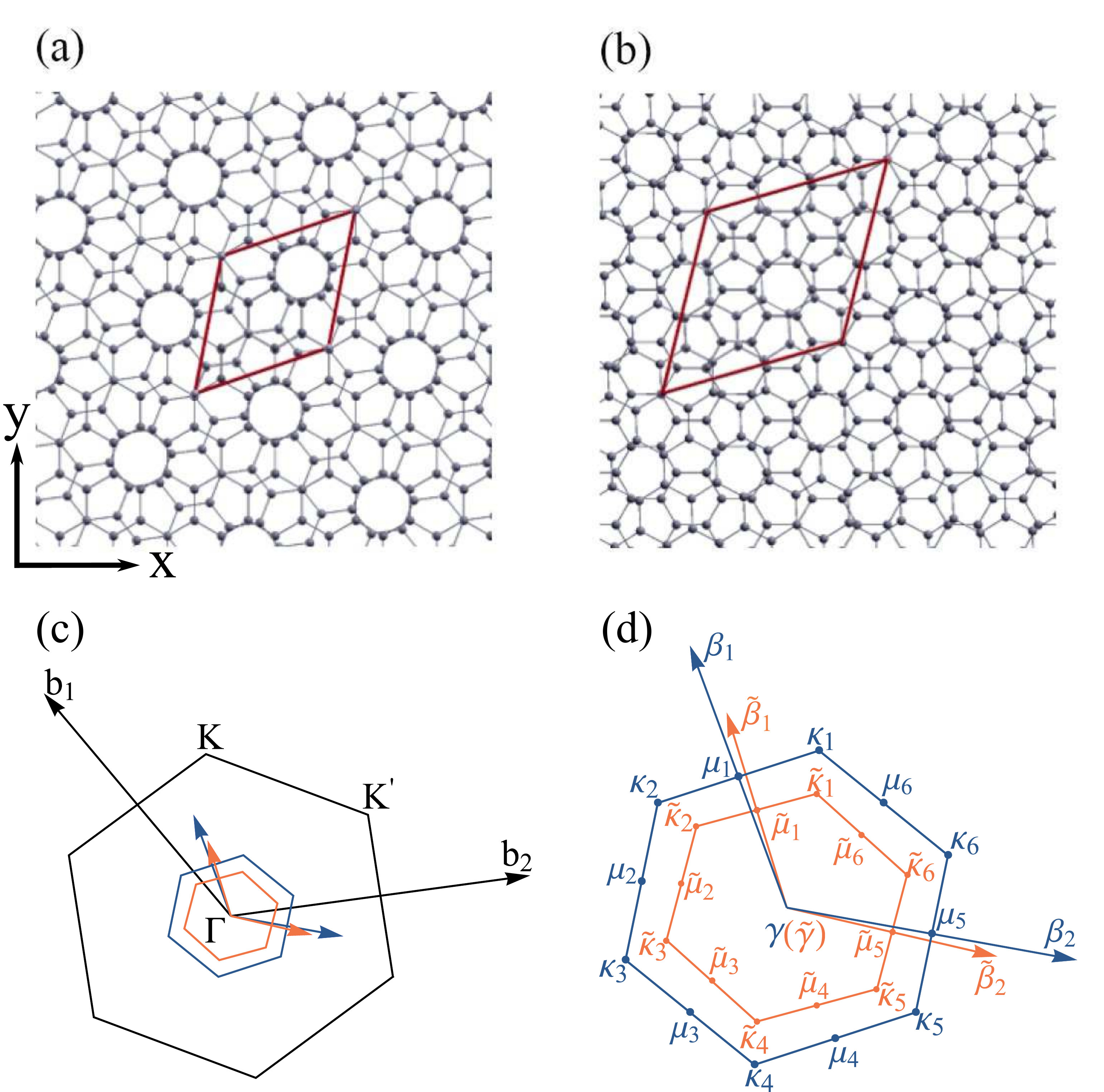}
\caption{(Color online). Top view of schematic representation of commensurate TBG (a) (1,4) with $\theta \approx 38.21^{\circ}$, (b) (1,3) with $\theta \approx 32.20^{\circ}$. Red parallelograms are the supercell of $(1,4)$ and $(1,3)$, each of which include 28 and 52 atoms, respectively. (c) First BZ and the reciprocal lattice vectors of graphene (black). Blue and orange hexagons are sBZ of $(1,4)$ and $(1,3)$ TBG, respectively. (d) Magnified sBZ of $(1,4)$ and $(1,3)$ TBG to aid the visualization of high symmetry points and namings. \label{fig1}}
\end{figure}
Here, to capture possible features driven by real space symmetries including the SE parity and investigate the relative electronic structure and their changes under applied in-plane strain we'll focus on two commensurate supercells, $(m,n)=(1,4)$ and $(1,3)$ that are even and odd under SE parity respectively (cf. Fig.\ref{fig1}). The real space superlattice and supercell Brillouin zone (sBZ) of (1,4) and (1,3) TBGs are depicted in Fig.\ref{fig1}. These structures have the shortest moir\'e pattern periodicity among the commensurate moir\'e structures and each of them consists of $28$ and $52$ atoms respectively. Also, the misalignment angle of the (1,4) and (1,3) superlattices, which are 38.21~\AA~and 32.20~\AA, have the smallest deviations from the twist angle of Bernal BG.  

To study the strain induced changes of the electronic properties of the (1,4) and (1,3) TBGs, we combine TB and first-principles calculations for both unstrained and strained TBGs. To this end, we perform first-principles calculations implemented in SIESTA code. We use double-$\zeta$ polarized basis (DZP) with Norm-conserving pseudopotential and the vdW exchange-correlation functional within the conjugate gradient method \cite{Soler2002}. Moreover, we sample the momentum space with $10\times 10 \times 1$ Monkhorst-Pack mesh grids. All DFT computations are converged over $400~{\rm Ry}$ energy mesh cutoff. The vacuum space perpendicular to the TBG layers is set to nearly $20~\AA$ to suppress the interactions between spurious images of the TBG. To obtain the band dispersion for unstrained TBG structures, we let both atomic coordinates and lattice vectors to relax until the forces on each atom become less than $0.04~{\rm eV/\AA}$.

To furthur study the low-energy physics of the TBGs performing TB method, we calculate the eigen energies and band dispersion of TBGs through the following Hamiltonian, 
\begin{eqnarray}
 \mathcal{H} = -\sum_{\langle i,j\rangle}
t(\Vec{r}_i - \Vec{r}_j) |\Vec{r}_i\rangle\langle\Vec{r}_j| + {\rm h.c.},
\label{eq_Hamiltonian_TBG}
\end{eqnarray}
where we use the tunneling integral equation \cite{Moon2014a},
\begin{eqnarray}
-t(\vec{d})=V_{pp\pi}(\vec{d})[1-(\frac{\vec{d}\cdot \vec{e}_{z}}{\vec{d}})^2]+V_{pp\sigma}(\vec{d})(\frac{\vec{d}\cdot \vec{e}_{z}}{\vec{d}})^2,
\label{tunn. int.}
\end{eqnarray}
to compute the hopping of the carriers in between single $p_z$ orbitals of carbon atoms located in graphene layers ($\Vec{r}_{i(j)}$) with relative distance $\vec{d}$. Here, we approximate the $a_0$ and $d_0$ with 1.42~\AA~and 3.3~\AA, which are the intra and interlayer distance between carbon atoms, respectively. Moreover, the $\pi$ and $\sigma$ hybridization energy of the $p_z$ orbitals are approximated by $V_{pp\pi}(\vec{d})=V^0_{pp\pi}exp(-(d-a_0)/\delta_0)$, $V_{pp\sigma}(\vec{d})=V^0_{pp\sigma}exp(-(d-d_0)/\delta_0)$. We choose the nearest couplings as $V^0_{pp\pi}=-2.7$ eV, $V^0_{pp\sigma}=0.48$ eV and the decay length constant as $\delta_0=0.184a_0$. 

Homogeneous lattice deformations which are uniform and equal in all in-plane directions, namely biaxial strains can be modeled by the change of the lattice constant. The more general lattice distortions that lead to the asymmetric deformation of the lattice structure including the uniaxial tensile strain can be modeled via changes of the lattice vectors. Here, to model the strained structures, we first modify the supercell vectors as, $\vec{R}_{i}^{\prime}=(1+\epsilon_{i})\vec{R}_{i}$, along any preferred direction $i=x,y$. Next, we optimize the TBG structure within DFT method by keeping the supercell vectors fixed at their strained values and letting the atoms to move. Within the TB approach, we use the same relation to alter the atomic coordinates to model the strained TBGs. Note that for both commensurate structures, based on our {\it ab initio} approach we find that the optimized positions of atoms deviate from the rigid ones used in TB method. However, for the sake of simplicity and also presenting a systematic approach applicable to any twist angle and external parameters such as strain, we use rigid atomic positions in the TB model.
\section{Results and discussions}\label{res}
%
%
%
%
\begin{figure}[!t] 
\includegraphics[width=\linewidth]{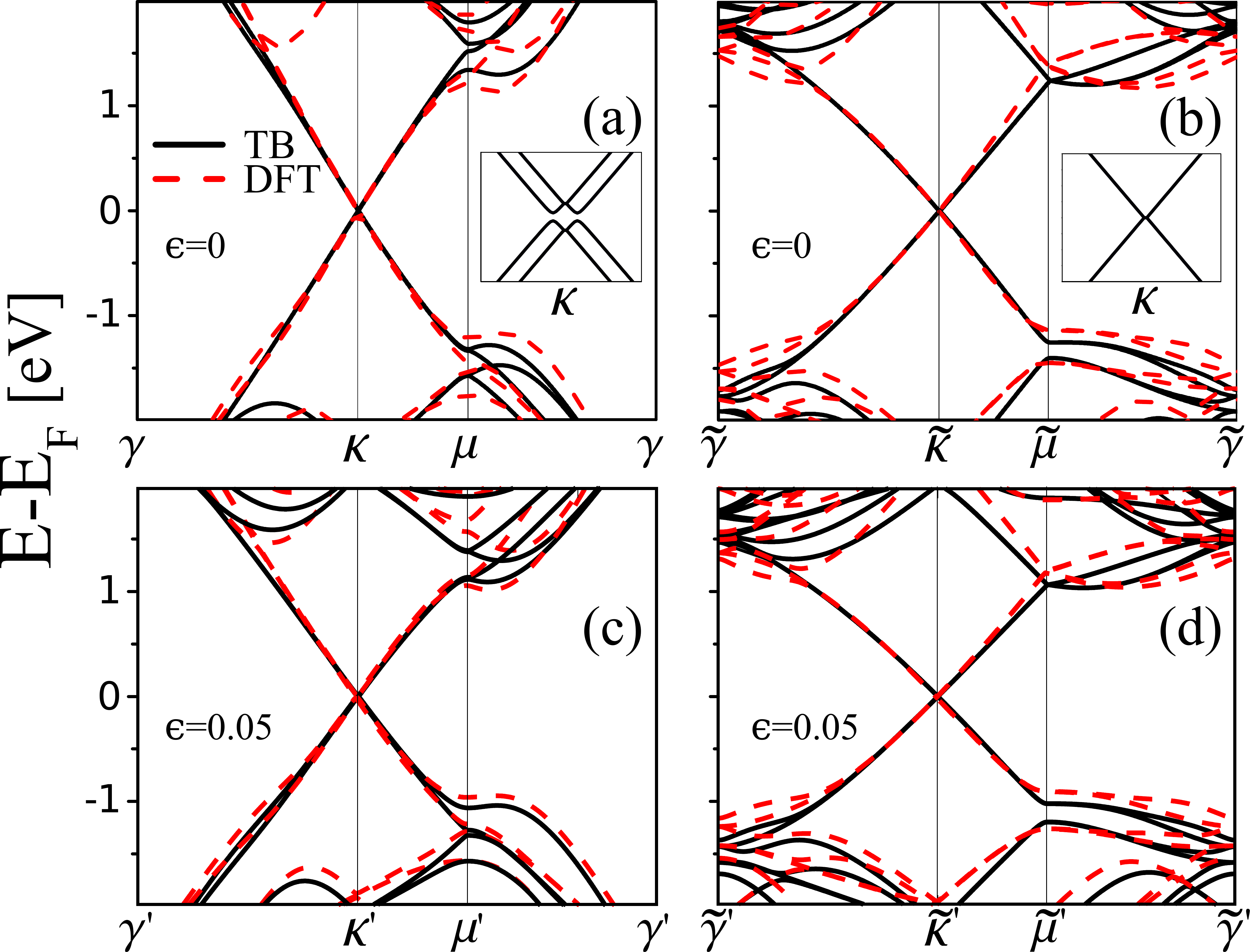}
\caption{Electronic band dispersion of (a,c) $(1,4)$ and (b,d) $(1,3)$ TBG along the high symmetry points of the first sBZ. While the lattice structures are unstrained in (a) and (b), the bands displayed in (c) and (d) belong to the commensurate TBGs when being exposed to 5\% biaxial strain. Red dashed (Black solid) curves corresponds to DFT (TB) results. TB and DFT calculated bands are in excellent agreement in high energy interval $[-1,1]~{\rm eV}$ both for unstrained and strained TBGs. Insets in (a) and (b) are zoomed-in figures of band structures in low-energy limit near Fermi energy where the energy interval is $[-20,20]~$meV. While the low-energy bands are massive and gapped for unstrained SE symmetric $(1,4)$, the bands of unstrained SE odd $(1,3)$ are gapless and linear in $\vec{k}$.
\label{fig2}}
\end{figure} 
Fig.\ref{fig2} illustrates the results for the DFT and TB computation of the band dispersion along the path of high symmetry points for TBG superstructures depicted in Fig.\ref{fig1}. The DFT and TB results are in good agreement and demonstrate Dirac fermionic behavior for both superstructures close to the charge neutrality point. In the insets where we show the electronic bands within a small energy interval of $[-20,20]$ meV, close to Dirac cone conical points, the low-energy features driven by the SE parity emerge. As it is evident from the insets, contrary to the gapless band dispersion of unstrained $(1,3)$ TBG, we clearly observe gapped massive Dirac cones for unstrained $(1,4)$ TBG when we zoom in the vicinity of charge neutrality point. The low-energy bands of the  $(1,3)$ are linear and degenerate in the scale of 20 meV and the massless Dirac cones intersect at the charge neutrality point. Hence, the energy interval in which the SE parity-driven low-energy behavior emerge scales inversely with the moir\'e period. Here our computations are consistent with the previous report of Ref \cite{Mele2012}. Also, our computations show that the renormalized Fermi velocity is 79 and 78 percent of the monolayer graphene Fermi velocity for the $(1,4)$ and $(1,3)$ superlattices respectively.

In panel (c) and (d), we present the band dispersion along the high symmetry points of the strained sBZ for $(1,4)$ and $(1,3)$ superlattices when applying 5\% biaxial tensile strain. The geometrical changes of the lattice structure relative to the applied strain lead to the modification of the lattice vectors, their dual space counterparts, and the sBZ. Within the TB approach, these changes affect the carrier hopping to the $p_z$ orbitals of carbon atoms through the modulation of the relative distances between the atoms (See Eq.\ref{tunn. int.}). The strain-induced changes of the DFT computed electronic structure of the TBGs which stem from the modification of expansion of the atomic orbitals and their hybridization, agrees well with those of the TB model. Furthermore, as it is can be seen from Fig.\ref{fig2}(c) and (d), despite the strong expansion of the lattice when applying 5\% biaxial strain both TBG superlattices retain their unstrained electronic structure and remain linear close to the charge neutrality point due to the fact that biaxial strain preserves the real space lateral symmetries. Moreover, we find that the Fermi velocity reduces to an almost equal value of 0.71 ${\rm v_F^0}$ for both superlattices with ${\rm v_F^0}$ being the graphene Fermi velocity. As a result, the applied biaxial strain flattens the band dispersion close to the Fermi energy.
%
%
%
%
\begin{figure}[!t]
\includegraphics[width=\linewidth]{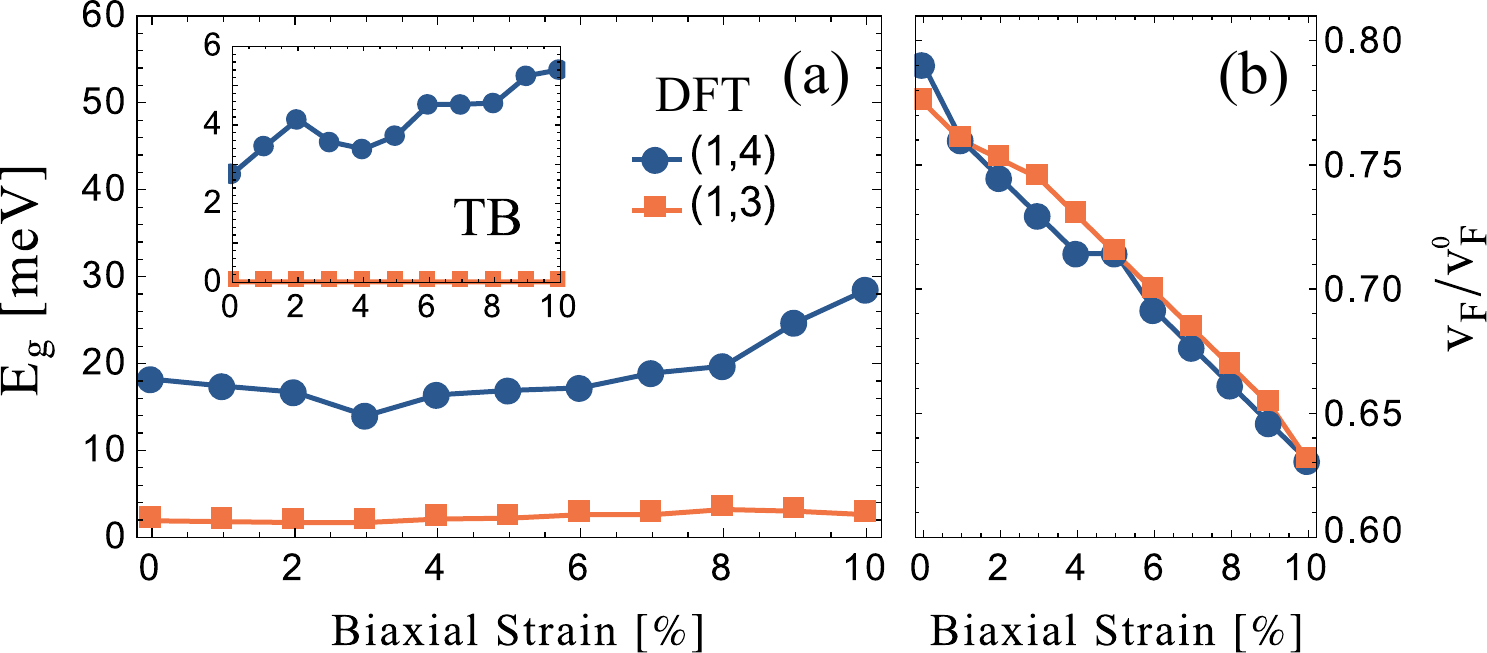}
\caption{(Color online). (a) DFT and TB (inset) calculated direct gap energy versus the applied biaxial strain on commensurate moir\'e structures shown in Fig.\ref{fig1}. The gap energy for the SE symmetric $(1,3)$ TBG is robust and remains unaffected by the applied biaxial strain, whereas the gap for the SE even $(1,4)$ TBG increases monotonically regarding the applied strain. The ratio of changes in gap energy calculated by DFT for the $(1,4)$ TBG is approximately 1 meV per percent of applied strain. (b) The Fermi velocity versus the applied biaxial strain for $(1,4)$ and $(1,3)$ superstructures. The renormalized Fermi velocity for both superstructures scales inversely with the applied biaxial strain. The strain-induced renormalization of the Fermi velocity is approximately 0.15 $\rm{v^0_F}$ per percent of applied strain. 
\label{fig3}}
\end{figure}

Now, we compute the modification of the band spacing and Fermi velocity when applying biaxial strain. Fig.\ref{fig3} shows the bandgap energy of commensurate $(1,4)$ and $(1,3)$ superlattices for tensile biaxial strains up to 10\% computed by both TB and DFT. Both methods represent qualitatively the same results as they show similar trends for the gap energy modulation regarding the applied biaxial strain. Interestingly, we see that the gap energy remains almost unchanged for the $(1,3)$ TBG even in presence of strong biaxial strains. Hence, the huge distortion of the lattice structure has a minor effect on the gap energy of the $(1,3)$ TBG. Furthermore, the applied in-plane biaxial strain is more efficient in the modulation of the gap energy for the $(1,4)$ superlattice and the rate of the changes is 1meV per percent of applied strain based on DFT. The relative difference between the reported values of the gap energy of TB and DFT stems from the lack of electron-electron repulsions in the TB approach. Furthermore, since we use optimized structures when computing the band dispersion within the DFT approach, instead of the rigid strained atomic positions, we effectively start with different lattice structures. Although these differences in the lattice structures are small, yet, they lead to a different expansion of the atomic orbitals and their overlaps.

In panel (b) we present the Fermi velocity as a function of the applied biaxial tensile strain for both TBG superlattices. The Fermi velocity is almost identical for both superstructures and reduces to 0.63 ${\rm v_F^0}$. Thus, the strain-induced renormalization of the Fermi velocity is approximately 0.15 $\rm{v^0_F}$ per percent of applied strain and the Fermi velocity for both superstructures scales inversely with the applied biaxial strain.

%
%
%
%
\begin{figure}[!t] 
\includegraphics[width=\linewidth]{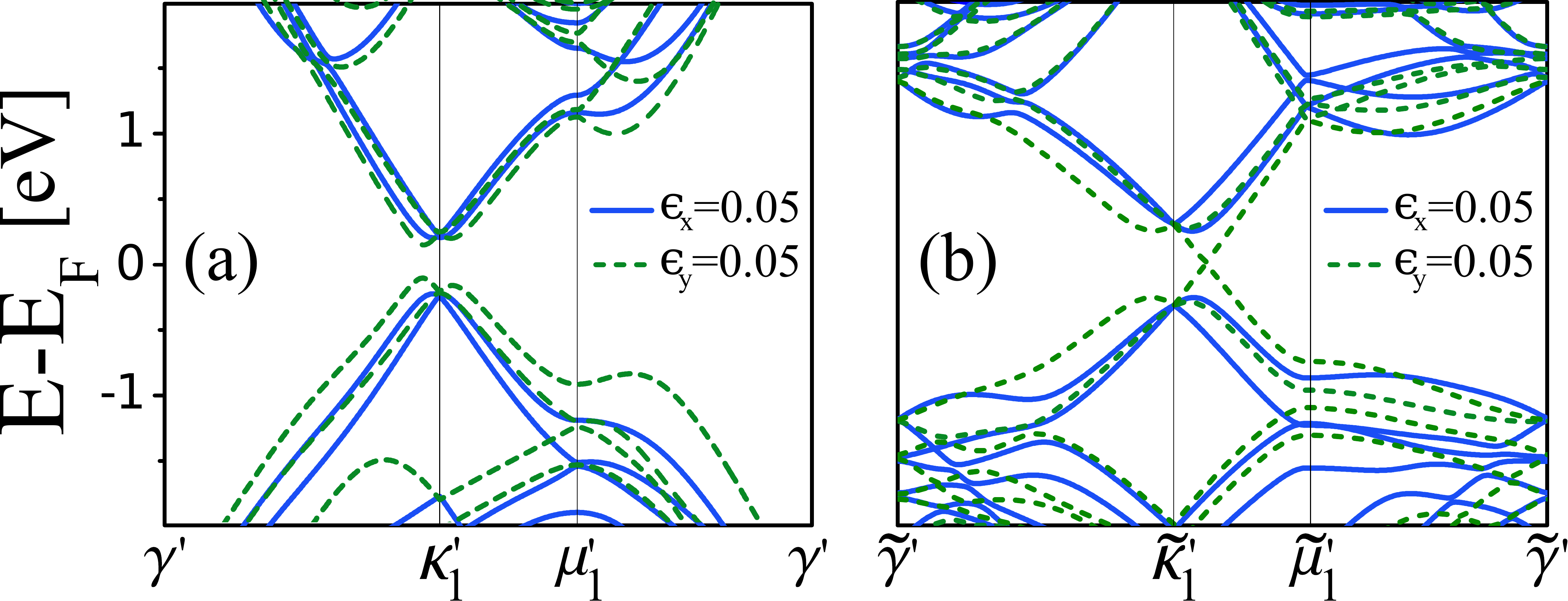}
\caption{Electronic band dispersion of strained (a) $(1,4)$ and (b) $(1,3)$ TBG along the high symmetry points of the first sBZ. Blue solid curves correspond to $5\%$ tensile strain along the $x$ axis and green dashed lines are for $5\%$ tensile strain imposition along the $y$ direction. The electronic dispersion is enormously altered after application of tensile strain for both TBG structures. All strained structures bear a huge gap energy close to K valley, except for strained $(1,3)$ in the $y$ direction where the bands stay linear and gapless and the Fermi energy single state dislocates from $\vec{k}=$K. Also, the electronic bands of $(1,3)$ become massive after imposition of $5\%$ strain along the x-axis.
\label{fig4}}
\end{figure}
We further compute the strain-induced modification of the gap energy regarding the biaxial strain for the next two smallest commensurate superstructures, $(1,7)$ and $(2,3)$ which are even and odd under SE parity. We find that similar to the $(1,3)$ superlattice, the band spacing is robust for the SE odd $(2,3)$ TBG and the band dispersion preserves its gapless behavior while being exposed to the biaxial strain. The gap energy modulation versus strain is shown in the Appendix. On the other hand, the band spacing increases monotonically with strain for the SE even $(1,7)$ superstructure, even though the gap is small. Moreover, the bandgap energy approaches 0 meV as the misalignment angle for SE even moir\'e superlattice becomes small in agreement with the previous report of Ref.\cite{LopesdosSantos2012}. Overall, when applying biaxial tensile strain, SE odd TBGs remain gapless and the SE even TBGs show increment of the gap energy.

Now to investigate whether the results of the biaxial straining, i.e. the SE odd $(1,3)$ TBG remains gapless and the gap energy for the SE even $(1,4)$ TBG increases with strain, can be generalized to other strain configurations, we study the non-equibiaxial and mixed strains. Fig.\ref{fig4} illustrates two exemplary strain configurations in which both $(1,4)$ and $(1,3)$ superstructures are exposed to uniaxial tensile strains (5\%) along the $x$ and $y$ directions. The asymmetric lattice distortion breaks the hexagonal symmetry of sBZ leading to three non-equivalent sBZ corners due to time reversal symmetry. Here we present the electronic bands for the strained TBGs along the path of high symmetry points close to $\kappa^{\prime}_1(\tilde{\kappa}^{\prime}_1)$. The SE even $(1,4)$ superstructure retains its unstrained electronic structure and stays gapped with two massive bands at the sBZ corner ($\kappa^{\prime}_1$). The gap energy, however, is enormously enhanced after applying uniaxial tensile strain in both directions. Compared to the unstrained $(1,4)$ superlattice the bands are flattened and the band velocity close to the charge neutrality point is reduced. Furthermore, strong band velocity discontinuity is observable for the uniaxially strained $(1,4)$ TBG along the $y$ axis at the $\kappa^{\prime}_1$. The $(1,3)$ bands close to the Fermi energy become massive after imposition of $5\%$ strain along the $x$ axis. All strained structures shown in Fig.\ref{fig4} possess a huge gap energy close to $\kappa^{\prime}_1$ point, except for uniaxially strained $(1,3)$ along the $y$ direction where the bands are gapless and the conical point of the Dirac cone drifts away from the sBZ corner ($\tilde{\kappa}^{\prime}_1$). Here, the low-energy bands follow two distinct features. One becomes massive and flattened at the sBZ corner and the other preserves the linear behavior of the unstrained Dirac fermions. Also, a strong band velocity discontinuity is observable at the sBZ corner.
%
%
%
%
%
\begin{figure}[!t]
\includegraphics[width=\linewidth]{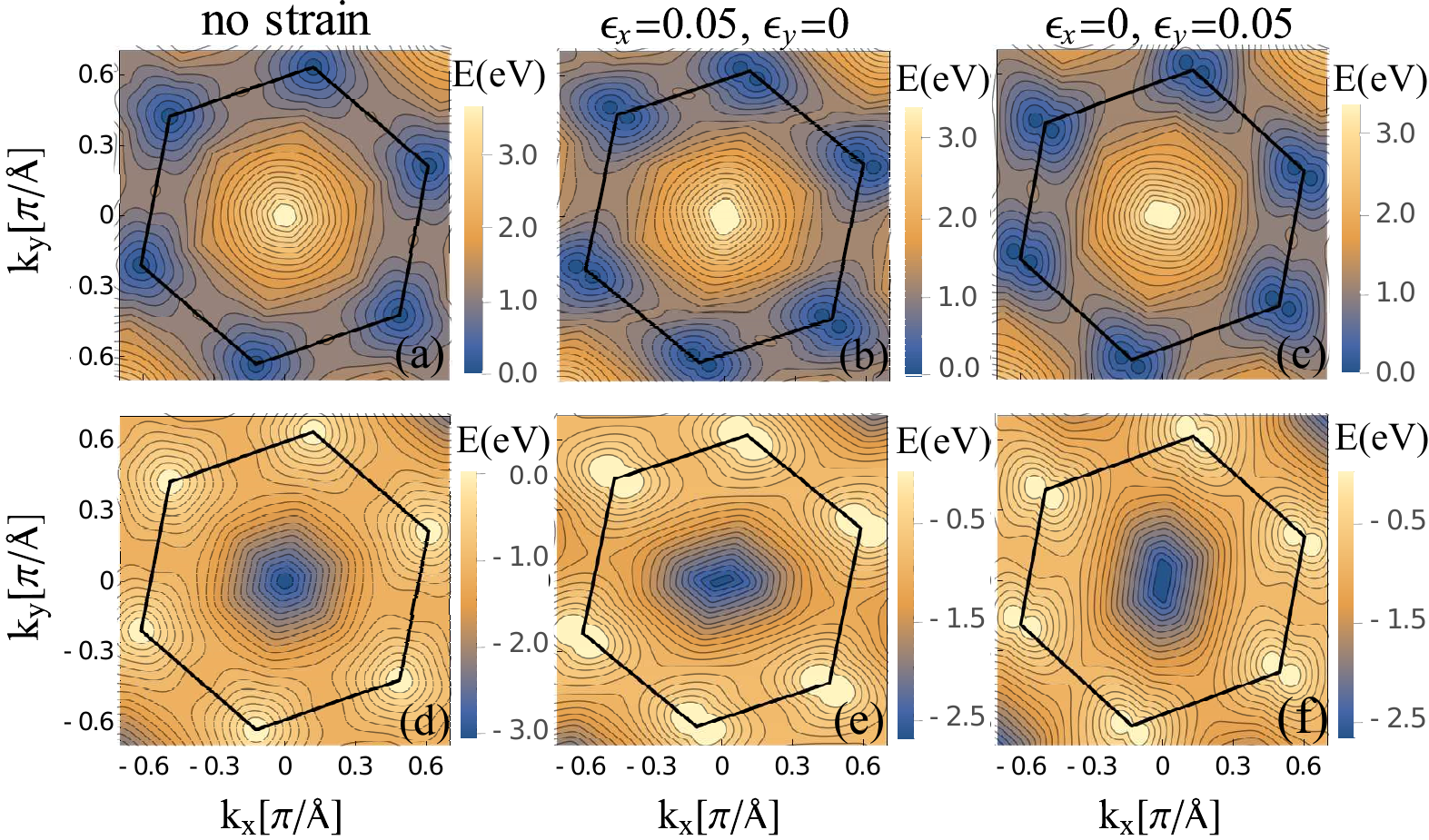}
\caption{(Color online). 
Surface maps of valence and conduction bands as a function of electronic wave vector for the diverse imposition of in-plane strain on $(1,4)$ TBG. Upper (lower) panels indicate conduction (valence) band. Black hexagons demonstrate the sBZ for the corresponding unstrained and strained structure. The band energies are shifted with regard to the Fermi energy so that the middle state between the highest occupied state and the lowest unoccupied state is set to 0 eV. Conduction band minimum and valence band maximum are displaced from the sBZ corners when applying strain, hence the gap energy shifts from sBZ corners. \label{fig5}}
\end{figure}

\begin{figure}[!t]
\includegraphics[width=\linewidth]{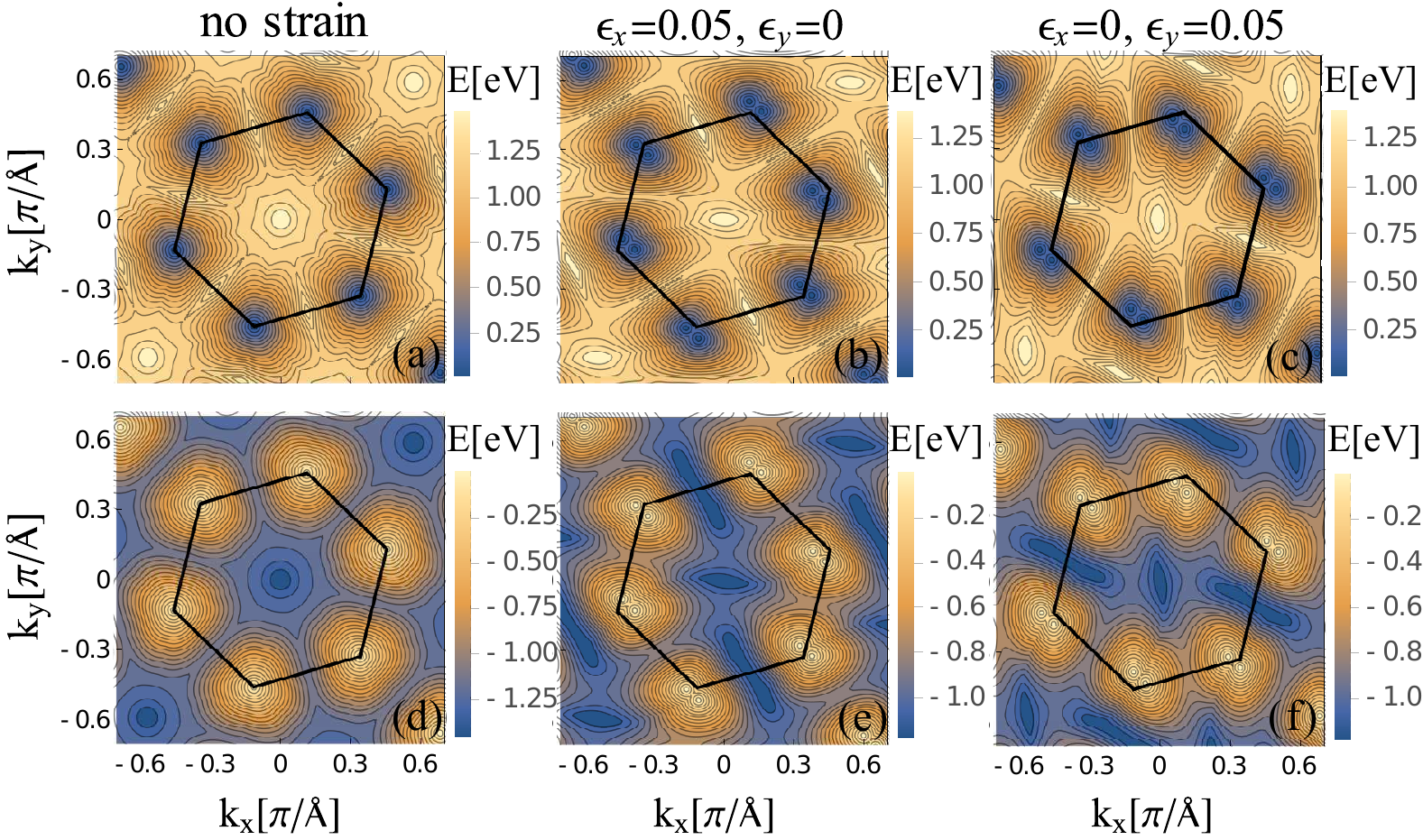}
\caption{(Color online). Same plots as in Fig.\ref{fig5} for TBG $(1,3)$. Analogous to the $(1,4)$ TBG, conduction band minimum and valence band maximum, and therefore the gap energy move away from sBZ corners after imposition of in-plane strain. \label{fig6}}
\end{figure}

To get a deeper understanding of the strained electronic bands and to evaluate the band spacing, we plot the surface maps of the lowest conduction and the highest valence band over the entire sBZ for unstrained $(1,4)$ and $(1,3)$ superlattices and the strained configurations displayed in Fig.\ref{fig4}. Fig.\ref{fig5} is the resolution of the low-energy bands for the $(1,4)$ TBG and the map plots of Fig.\ref{fig6} are those of the $(1,3)$ superlattice. All electronic bands displayed in figures \ref{fig5} and \ref{fig6} are shifted regarding the undoped Fermi energy state so that the middle state between the highest occupied level and the lowest unoccupied level is 0 eV. The first row of both plots shows the lowest conduction band whilst the second rows are the highest valence bands of the corresponding strain configuration. Moreover, Black hexagons depict the sBZ of the corresponding lattice structure. We clearly observe trigonal warping due to the non-orthogonal interlayer couplings \cite{Abergel2010}, close to the sBZ corners of both valence and conduction bands for unstrained TBGs. This sublattice broken symmetry driven by interlayer interactions also results in the renormalization of the Fermi velocity of TBGs observable in Fig.\ref{fig2}(a,b). The threefold anisotropic behavior of the Fermi lines is strongly distorted as the uniaxial tensile strain is applied on both TBG superstructures. Therefore, the isoenergy lines at the sBZ corners for the unstrained TBGs split into two observable isoenergy pockets leading to the relocation of the conical point of the Dirac cone minibands from the sBZ corners. Thus, similar to the case of uniaxially strained monolayer graphene that the Dirac cones and hence the single Fermi state dislocate from the BZ corners \cite{Pereira2009a}, the real bandgap energy of the uniaxially strained TBGs locates beyond the sBZ hexagon and cannot be identified along the path of the high symmetry points. Consequently, the bandgap should be evaluated with care. Furthermore, the strain-induced changes of the trigonal warping driven by modified interlayer coupling when applying uniaxial tensile strain are responsible for the reduction of the Fermi velocity and flattening of the bands in both TBG superstructures (cf. Fig.\ref{fig4}). Also, the anisotropic strain-induced distortion of the Fermi lines results in the band velocity discontinuity observable at sBZ corners in Fig.\ref{fig4}. Our computations reveal that the bandgap energy for the commensurate $(1,3)$ superstructure when stretched along the $x$ ($y$) axis with 5\% uniaxial strain is 3 meV (4 meV) as a consequence of the broken real space symmetry. Therefore, the uniaxial tensile strains lead to bandgap opening in the SE odd $(1,3)$ superlattice. 
%
%
%

\begin{figure}[t!]
\includegraphics[width=\linewidth]{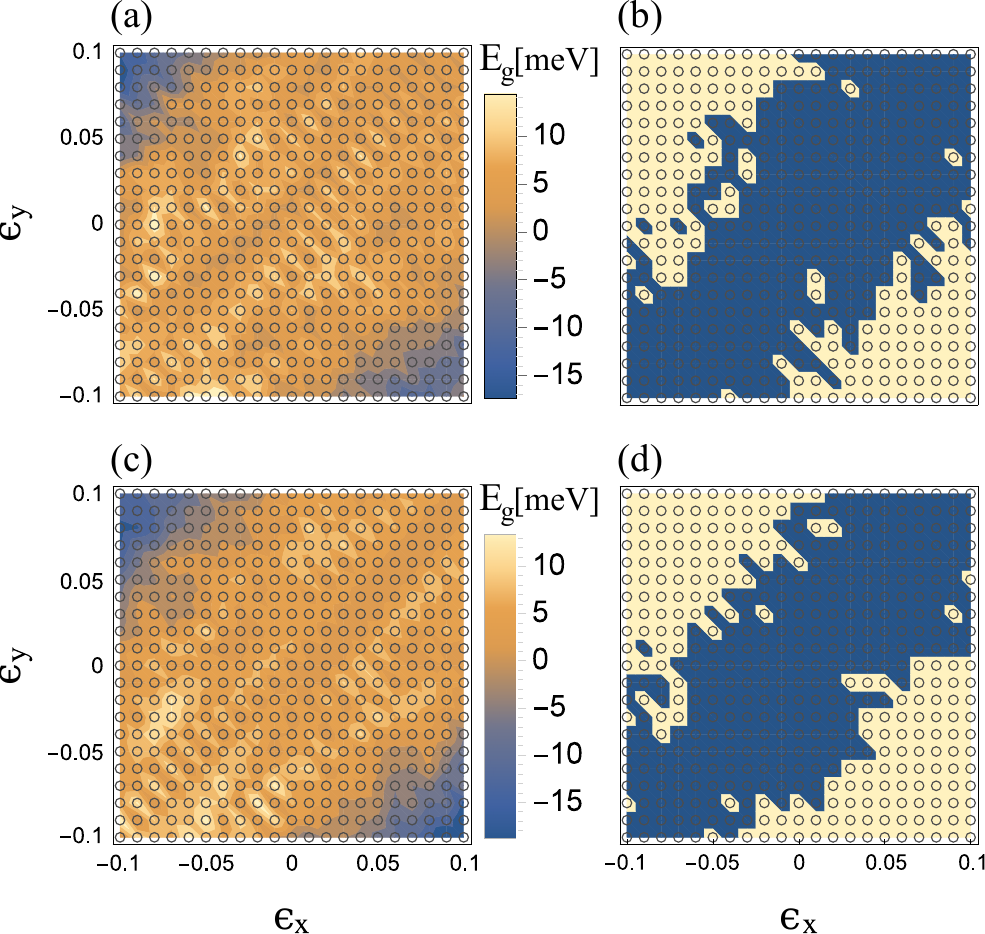}
\caption{(Color online). TB calculated surface plots of bandgap energy as a function of applied strain on (a) $(1,4)$ TBG and (c) $(1,3)$ TBG. The empty small circles correspond to real data and the background is the fitted formula. The resolution of the bandgap as direct or indirect for (b) $(1,4)$ and (d) $(1,3)$ TBGs. Bright areas in (b) and (d) demonstrate strain configurations in which the bandgap of the corresponding TBG is indirect. For highly mixed strain configurations in which the system is stretched along one in-plane direction and compressed along the other in-plane direction, the bandgap for both TBG structures becomes indirect. Non-equibiaxial compressive strains are more efficient in increasing the direct gap for both commensurate TBG structures. \label{fig7}}
 \end{figure}

Now, we use the TB model as a computationally less expensive method to investigate the modification of the bandgap with respect to the applied in-plane strain. Fig.\ref{fig7} is the resolution of the gap energy for diverse strain configurations of the $(1,4)$ and $(1,3)$ superlattices. The gap energy is calculated for 441 strain configurations of the commensurate TBGs which are depicted as empty circles in Fig.\ref{fig7}. These configurations include biaxial, uniaxial, compressive and mixed strains where the absolute value of the strain along the in-plane directions increases up to 10\%. Panels (a) and (c) are the band spacing (direct and indirect) for $(1,4)$ and $(1,3)$ superlattices, respectively. Here we diagonalize the TB Hamiltonian over a dense mesh grid of wave vector ($\vec{k}$) in reciprocal space. Next we define and evaluate the least band spacing as ${\rm min(E_c}(\vec{k}))-{\rm max(E_v}(\vec{k}))$. Fig.\ref{fig7}(b) and (d) indicate the type of the bandgap, that is the bright areas depict the indirect and the dark blue areas show the direct bandgaps for the corresponding strain configuration. Interestingly, the bandgap for both commensurate TBG structures becomes indirect when applying strong mixed strains (cf. top left and bottom right corner of the panel (a) and (c)).
%
%
%
%
%
%
%
%
Moreover, the valence band maximum becomes energetically higher than the conduction band minima but in different valleys when both TBG structures are exposed to extremely large mixed strains (cf. the strain configuration $\epsilon_x=0.1$ and $\epsilon_y=-0.1$). Note that our DFT computations show that for the extremely large strained structures where the system is stretched along one in-plane direction and compressed along the other in-plane direction, the system remains integrated and the strained TBG is in the elastic region. As can be seen from the panel (a) and (c), the maximum value of gap energy is observable in symmetry broken highly compressed structures, generally when the compressive strain is along both directions but with different magnitude (See the bottom left corner of Fig.\ref{fig7}(c)). There also exist some mixed strain configurations that are efficient in increasing of the bandgap for $(1,4)$ and opening of the bandgap for $(1,3)$ TBG (cf. the strain situation $\epsilon_x=-0.05$ and $\epsilon_y=0.01$ for $(1,4)$ and the $\epsilon_x=-0.05$ and $\epsilon_y=0.02$ situation for the $(1,3)$). For both TBG structures studied here the bandgap does not exceed 15 meV. The general trend for the modification of the bandgap of the $(1,4)$ and $(1,3)$ superlattices regarding the in-plane strain is similar, except for the biaxial and some specific strain configurations. Contrary to the biaxial strain where the $(1,3)$ TBG remains gapless and unaffected by the lattice deformations, other strain configurations result in the emergence of a bandgap. Moreover, the equibiaxial compressive strain is not efficient in gap opening for $(1,3)$ TBG. In fact, for SE odd $(1,3)$ TBG when the lattice distortions are equal in all in-plane directions and the symmetries are not broken, the bandgap is 0 eV. 
\section{Conclusion}\label{concl}
We have studied the impacts of in-plane strain on electronic properties of two exemplary SE odd and SE even commensurate TBG superlattices with large twist angle. We observed that the biaxial tensile strains leave the low-energy behavior of the SE odd TBG, i.e the gapless massless Dirac cones at the neutrality point, unchanged whilst they lead to an increment of bandgap for the SE even superlattice. Furthermore, we found that the renormalized Fermi velocity for both superstructures scales inversely with the applied biaxial strain. We took the advantage of the reasonable agreement between the TB and DFT calculated band dispersion of the unstrained and strained TBGs to tackle more than 400 strain configurations via the less computationally expensive approach of TB. There we found that for specific mixed strains where the TBGs are stretched along one in-plane direction and compressed along the other in-plane direction, both superlattices show direct-indirect bandgap transition. Consequently, the large angled commensurate TBGs are promising platform for manipulation of the electronic structure at low strain costs, specifically because the fabrication of the moir\'e structures with controlled stacking is experimentally feasible. 
\section*{ACKNOWLEDGEMENTS}
We thank David S. L. Abergel for helpful discussions.
\section*{APPENDIX}
Similar to the $(1,3)$ TBG superstructure the gap energy of the SE odd $(2,3)$ superstructure is unaffected by the biaxial tensile strain even when the strain is strong. On the other hand, the in-plane biaxial strain leads to the increment of the gap energy for the SE even $(1,7)$ superlattice analogous to the $(1,4)$ TBG. The rate of the changes in the gap energy is 0.1 meV/\%.
\begin{figure}[!htbp]
\includegraphics[width=.7\linewidth]{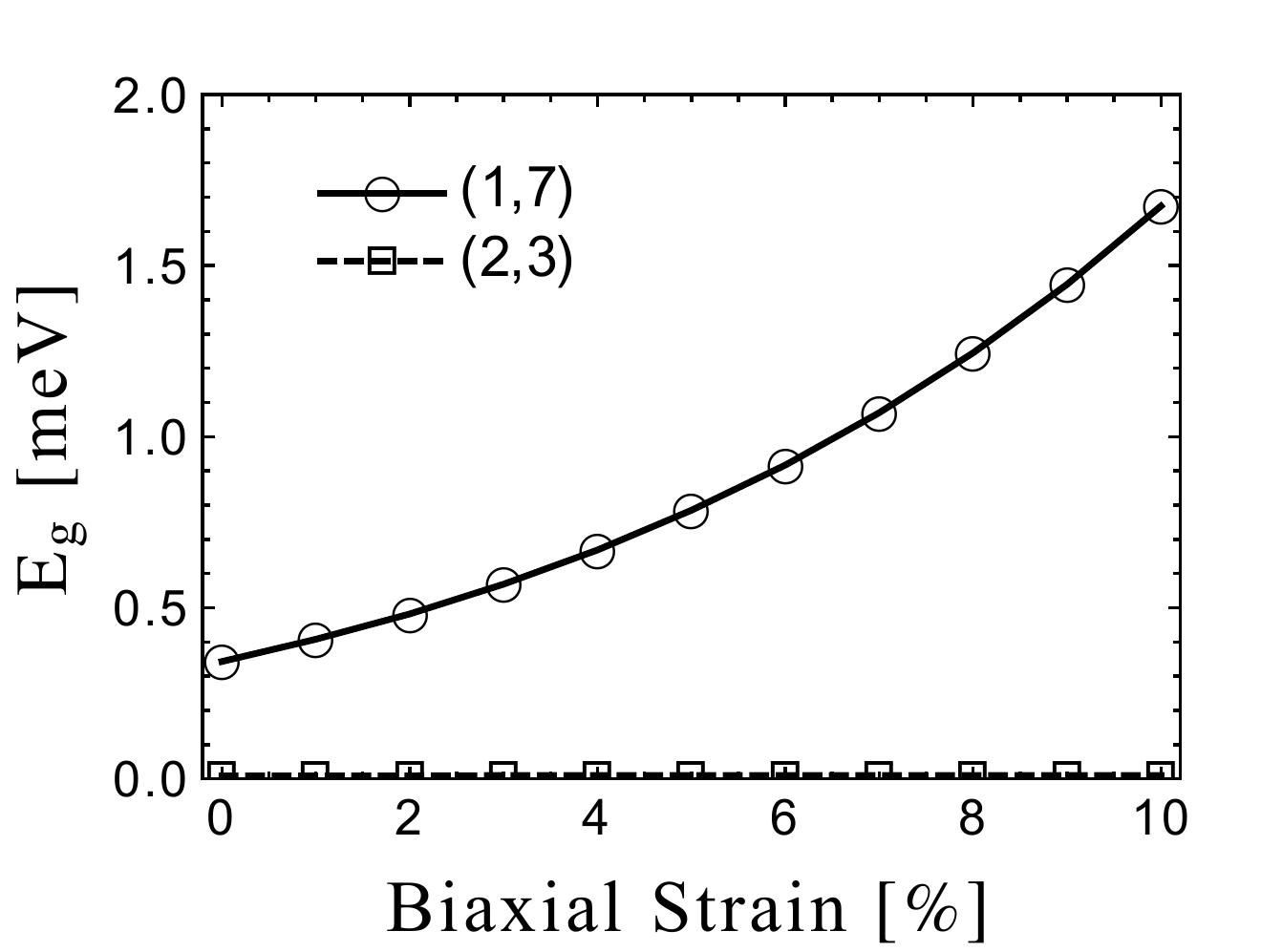}
\caption{(Color online) Gap energy versus the applied biaxial strain for $(1,7)$ and $(2,3)$ moir\'e superstructures. The SE odd $(2,3)$ TBG remains gapless when applying biaxial tensile strain, whereas the gap for the SE even $(1,7)$ superlattice scales monotonically with the applied strain. \label{a1}}
\end{figure}

\section*{REFERENCES}
\bibliography{ref.bib}
\end{document}